\documentclass[acmsmall,screen,manuscript]{acmart}
\usepackage{multirow}
\usepackage{xspace}
\usepackage{color}

\newcommand{\revised}{\color{black}}

\AtBeginDocument{%
  \providecommand\BibTeX{{%
    \normalfont B\kern-0.5em{\scshape i\kern-0.25em b}\kern-0.8em\TeX}}}

\setcopyright{acmcopyright}
\copyrightyear{0000}
\acmYear{0000}
\acmDOI{00}

\acmJournal{JETC}
\acmVolume{0}
\acmNumber{0}
\acmArticle{0}
\acmMonth{0}

\begin{document}

\title{Image Complexity Guided Network Compression for Biomedical Image Segmentation}


\author{Suraj Mishra}
\email{smishra3@nd.edu}
\orcid{1234-5678-9012}
\author{Danny Z. Chen}
\email{dchen@nd.edu}
\author{X. Sharon Hu}
\email{shu@nd.edu}
\affiliation{%
  \institution{University of Notre Dame}
  \streetaddress{Department of Computer Science and Engineering}
  \city{Notre Dame}
  \state{Indiana}
  \country{USA}
  \postcode{46556}
}

\begin{abstract}
Compression is a standard procedure for making convolutional neural networks (CNNs) adhere to some specific computing resource constraints. However, searching for a compressed architecture typically involves a series of time-consuming training/validation experiments to determine a good compromise between network size and performance accuracy. To address this, we propose an image complexity-guided network compression technique for biomedical image segmentation. Given any resource constraints, our framework utilizes data complexity and network architecture to quickly estimate a compressed model which does not require network training. Specifically, we \textit{map} the dataset complexity to the target network accuracy degradation caused by compression. Such mapping enables us to predict the final accuracy for different network sizes, based on the computed dataset complexity. Thus, one may choose a solution that meets both the network size and segmentation accuracy requirements. Finally, the mapping is used to determine the convolutional layer-wise multiplicative factor for generating a compressed network. We conduct experiments using 5 datasets, employing 3 commonly-used CNN architectures for biomedical image segmentation as representative networks. Our proposed framework is shown to be effective for generating compressed segmentation networks, retaining up to $\approx 95\%$ of the full-sized network segmentation accuracy, and at the same time, utilizing $\approx 32x$ fewer network trainable weights (average reduction) of the full-sized networks. 
\end{abstract}



\maketitle

\section{Introduction}
Biomedical image segmentation plays a key role in disease diagnosis and treatment. Recently, by outperforming traditional approaches, convolutional neural networks (CNNs) have become powerful tools for biomedical image segmentation. In one such CNN based early work, Ronneberger~\textit{et al.}~\cite{unet} achieved state-of-the-art accuracy in segmenting neuronal structures by proposing U-Net. Since its inception, U-Net has become one of the most popular CNN models for biomedical image segmentation. Networks like CUMedVision~\cite{cumednet}, coarse-to-fine stacked networks~\cite{coarsetofine}, cascaded networks~\cite{cascaded}, U-Net++~\cite{zhou_r}, and UCU-Net~\cite{mishra} were also designed to improve biomedical image segmentation accuracy. Such networks outperform traditional methods and are considered currently as state-of-the-art for many tasks such as melanoma segmentation~\cite{mel_winner, mel_320, me_cvpr_data}, lymph node segmentation~\cite{yizhe}, and retinal vessel segmentation~\cite{wu_new,mishra,mou,yishuo,uncert}. However, CNNs are often of very large sizes, resulting in high memory requirements and high latency of operations, and thus may not be suitable for resource-constrained applications (e.g., edge computing).  

Nowadays, low cost and easy-to-carry (e.g., handheld) imaging devices are widely used in edge computing type of biomedical and healthcare applications (e.g., disaster/emergency response, pandemic management, and military rescue), and desirably, the most effective image analysis techniques, including deep learning methods, are applied. However, in many edge computing scenarios (e.g., in {\revised remote or resource constrained areas}, battlefields, etc), computing resources may be severely limited and cannot implement the ordinary (full) deep learning network models. Hence, compressed versions of deep learning models, subject to local computing resource constraints, should be deployed to achieve best possible performance.

Neural network compression is an important aspect of neural network design. Benefits of compression include faster training, faster inference, and less resources required to design more energy-efficient applications. Post-training compression techniques 
such as pruning (removing less important filters) and quantization (using lower-precision representations for weights) have been proposed~\cite{deepcompression,nvidiapruning,qbert,incremental_quant,ch_split_quant}. Pre-training compression approaches focus on designing smaller networks to begin with~\cite{squeezenet,mobilenet}. Although these techniques are quite effective in finding smaller networks with acceptable accuracy, they require some parameters to be set manually and use multiple pruning -- fine-tuning iterations. In most cases, one standardized big network for segmentation is used regardless of the input data. Hence, compression often commences with the same large initial network and incurs lots of computation overhead. Howard~\textit{et al.}~\cite{mobilenet} proposed to reduce network size using a uniform multiplicative factor for each convolutional layer, which can quickly produce a smaller network. However, no systematic approach was provided to determine the value of the multiplicative factor. Hence, searching for a compressed CNN architecture for a specific imaging application using~\cite{mobilenet} typically involves a series of time-consuming training/validation experiments using the training data to find a good compromise between network size and performance accuracy. Further, a uniform multiplier based approach is not effective as different convolutional layers in a CNN do not contribute equally to feature extraction~\cite{raghu}. To address these challenges, in this paper, we propose a layer-wise multiplier based network compression framework targeting biomedical image segmentation in resource-constrained application settings, which quickly estimates a compressed model by exploiting properties inherent to the target application datasets.

For biomedical image segmentation, depending on the specific diseases or biological targets, the application datasets often exhibit distinctive properties that may shed light on how large of a network may be needed for segmenting the corresponding images. In contrast to natural scene images, in biomedical/healthcare application (or some application-specific) settings, images are often for a specific type of disease/injury and captured by specific imaging devices; hence, their objects and settings are quite ``stable'', making the image characteristics and complexity much easier to analyze. We leverage this useful property of biomedical images and propose to use image complexity as a guide to analyze segmentation accuracy degradation caused by compression.

Compressing a CNN by removing network weights \textit{generally} results in accuracy degradation. It is intuitive that a compressed network may not be able to capture robust image features well with fewer resources (i.e., fewer trainable network weights). We hypothesize that the drop in segmentation accuracy of a CNN caused by compression follows a pattern that can be linked to the target dataset complexity. This assumption is coherent with information theory as we believe \textit{`less'} complex images contain fewer features and hence can be captured by fewer network weights (or can be compressed more) while \textit{`more'} complex images require a larger amount of network weights to be successfully captured. Hence, compressing by pruning network weights will have different accuracy degradation on the same network for two different image datasets with different complexities. We seek to map this relation between dataset complexity and network accuracy degradation and call it \textit{degree of degradation}. We believe that for a network architecture, its \textit{degree of degradation} is a constant and can be estimated by tracking the accuracy degradation with network compression. Once calculated, the \textit{degree of degradation} can be utilized with the dataset complexity to predict accuracy degradation on any target dataset that will be caused by network compression.   

In this paper, we introduce a new framework for efficiently producing low latency and compressed deep learning networks for biomedical image segmentation without repeated training. We exploit the concept of training data complexity to guide the design of the compressed networks. Specifically, we quantify the complexity of the training image dataset and use it as an indicator of the target network's trainable weight requirements. We propose several complexity metrics for this purpose, which are much less computationally demanding than CNN training. Then, we map the calculated image complexity with the accuracy degradation of the CNN caused by compression to extract the \textit{degree of degradation} information. Using the computed image complexity of the training dataset and the \textit{degree of degradation} of the target architecture, we predict the accuracy for different network sizes without conducting network training. Thus, one may choose a solution that meets both the size and accuracy requirements. Based on the complexity measure, the target network architecture, and specified network constraints (e.g., accuracy or available memory), we determine the most suitable layer-wise multiplicative factors for the given dataset that translates to a compressed network. The resulting compressed network is then trained {\revised from scratch}, with much less effort and memory compared to a full network for image segmentation. Our approach complements post-training network reduction techniques, by focusing on the pre-training stage to quickly generate a size-reduced network structure for training. We conduct experiments using 3 publicly available and 2 in-house datasets, employing 3 commonly-used CNN architectures for biomedical image segmentation as representative networks to highlight the efficacy of our proposed framework.

Our main contributions are as follows: 
\begin{itemize}
\item Introducing a novel approach for compressing target CNNs for biomedical image segmentation based on image complexity, network architecture, and design constraints.

\item Analyzing various measures for representing image complexity and their suitability for guiding network compression.

\item Validating our approach on 3 representative biomedical image segmentation networks to generate corresponding compressed network architectures. 
\end{itemize}

\begin{figure}[tb]
    \centering
    \includegraphics[width=0.8\textwidth]{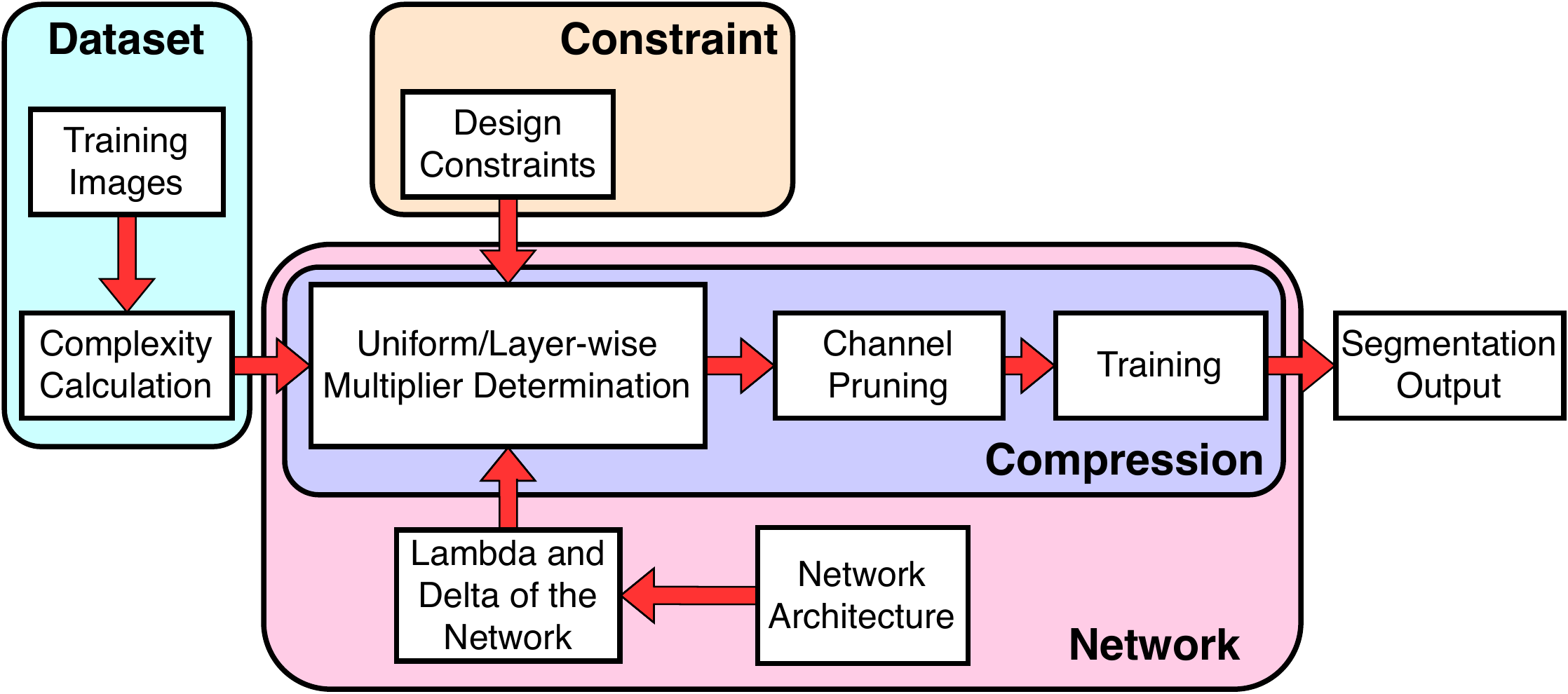}
    \caption{Our proposed framework for design constraint based network compression. Given an allowed network architecture degree of degradation and the image complexity of a target dataset, we compute the most suitable multiplier based on the design constraint. Then the compressed network generated using the multipliers is trained. {\revised Target dataset, design constraint, and network architecture are the required inputs for our framework (shown in cyan, brown, and magenta boxes, respectively). The degree of degradation associated with the network architecture is represented as two network-specific parameters lambda and delta (see Section~\ref{sec:net-param}).}}
    \label{fig:net-comp}
\end{figure}

Our proposed framework (shown in Fig.~\ref{fig:net-comp}) has three major components: (1) image complexity calculation, (2) network degree of degradation calculation, and (3) design constraint inclusion. In Section~\ref{sec:complexity}, we provide the details of image complexity calculation. In Section~\ref{sec:net-param}, degree of degradation calculation for neural networks is presented. Using user specified constraints to explore the design space is described in Section~\ref{sec:sec-user-const}. Experimental evaluations and discussions are provided in Section~\ref{sec:exp} and Section~\ref{sec:discuss}, respectively. Section~\ref{sec:conclusion} concludes the paper.

\section{Image Complexity Computation}
\label{sec:complexity}
In this section, we first explore various candidates for measuring image complexity in Section~\ref{ssec:m_comp_cand}. Then we present our approach to select the target complexity measure in Section~\ref{ssec:m_comp-select}. Finally, we propose a method to compute the layer-wise image complexity which will enable us to perform fine-grain layer-wise pruning.

\subsection{Complexity Candidates}
\label{ssec:m_comp_cand}
Our goal of exploring various image complexity measures is to identify an indicator that represents the information content of data samples. We seek an image complexity metric that can (i) indicate the trend of segmentation accuracy and (ii) be easily computed. Our work examines the following candidate metrics for image complexity estimation.

{\textbf{Signal Energy:}}
The summation across all squared coefficients of the frequency spectrum of a signal is taken as the signal energy. In CNN, we essentially perform filtering of various spatial frequency components present in the images. Hence, higher energy can be attributed to the presence of a richer frequency spectrum, and this may be considered as an indicator for a larger number of filter kernels in CNN to extract valuable information from the data. Similarly, lower energy of an image can be translated to a need for a smaller number of filters in CNN. To compute the image energy of a single image, we calculate the sum of the squared absolute values of the Fourier coefficients.

{\textbf{Edge Information:}}
Since segmentation focuses on detecting the boundaries of the objects of interest, we consider edge information as an important component of image complexity estimation. Yu and Winkler \cite{imagecomplexity} used edge information to calculate image complexity. Spatial information at the pixel level is calculated by summing the squared horizontal and vertical edge information extracted using horizontal and vertical Sobel or Scharr kernels, respectively. We compute edge information at different scales to imitate CNN-based fine-to-coarse feature extraction. The mean value of the edge information at different levels is used as an indicator of the image complexity.

{\textbf{Local Keypoint Detection:}}
Traditional local keypoint extraction approaches, such as SIFT \cite{sift} and SURF \cite{surf}, are widely used for computer vision tasks. We consider the number of extracted SURF keypoints, along with their strengths, as another estimate of image complexity.

{\textbf{Visual Clutter:}}
Rosenholtz {\it et al.}~\cite{visualclutter} presented a study of visual clutter, and its effect on feature extraction was provided. The presence of clutter affects visual tasks since it makes feature extraction more complicate. Hence, clutter can serve as a candidate for image complexity estimate. We consider feature congestion and sub-band entropy clutter measures for complexity computation. Feature congestion represents a subjective interpretation of visual clutter, while sub-band entropy is related to the visual information on the display.

{\textbf{JPEG Compression:}}
JPEG-based complexity utilizes a JPEG image compressor. The JPEG-based complexity is defined as the inverse of the compression ratio, {\it i.e.}, $\frac{1}{CR}$, where
\begin{equation}
    CR = \frac{size(image)}{size(compressed\_image)}.
    \label{eqn:jpeg}
\end{equation}
 The compressed image is generated using JPEG compression at 25\% quality \cite{imagecomplexity}. A higher JPEG complexity represents a less compressed image with less redundant information. A lower JPEG complexity signifies the presence of redundant information with a higher compression ratio.

{\textbf{Foreground Density:}}
The foreground density accurately represents correlation between foreground and background pixels and can be easily computed as a ratio of the number of foreground pixels to the number of the total pixels in an image, i.e., $B = \sum_{i} fg\_pixel / \sum_{i} img\_pixel$.

\begin{figure*}[tb]
    \centering
    \includegraphics[width=0.65\textwidth]{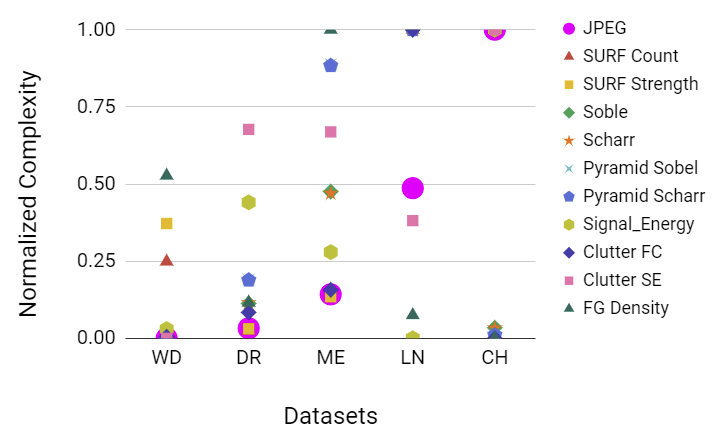}
    \caption{Mapping image complexity with accuracy degradation. In X-axis, datasets are arranged in increasing order of drop in F1 score with network compression. }
    \label{fig:complexity_trend}
\end{figure*}

\subsection{Candidate Selection}
\label{ssec:m_comp-select}

Given that there are multiple training images in a training dataset, we use the mean of the complexity values of all the training images for a specific measure as the corresponding complexity value. Since biomedical images for a specific application (e.g., a specific disease or injury) are often captured by the same imaging modality and contain fixed types of objects, it is reasonable to expect a relatively small variation among the complexity values of different image samples in the same dataset (if an appropriate complexity measure is used). The average complexity value of the training data can then be considered as the representative complexity of the image data for that application.

To see which of the above complexity measures is the most suitable to be used as a guide to direct network compression, we map these complexity measures to segmentation accuracy drop during network compression (to be explained in Section~\ref{sec:net-param}). Since F1 score is the most suitable and robust metric for capturing accuracy in class imbalance problems along with being one of the most used accuracy metrics, we explore F1 score for measuring segmentation accuracy in our framework. In Fig.~\ref{fig:complexity_trend}, different complexity measures (min-max normalized) are plotted against the F1 score degradation. The trends are different for most of the complexity  measures. Compared to other complexity measures, the JPEG complexity clearly follows the trend of F1 score degradation, i.e., higher JPEG complexity values lead to higher F1 score degradation with compression, as shown in Fig.~\ref{fig:complexity_trend}.
 
Along with F1 score, meanIU or IU (class-wise mean of Intersection over Union) is another commonly used metric~\cite{coarsetofine} to measure segmentation accuracy. Since IU relates to both feature variety and quantity, besides JPEG complexity, we introduce a new complexity measure which combines the JPEG complexity and foreground density, denoted by JB. Specifically, JB is defined as a linear function of the JPEG complexity and foreground density, i.e., $JB = \omega J + (1 - \omega) B$, where $J$ is the JPEG complexity, $B$ is the foreground density, and $\omega$ is a value in $[0,1]$. The value of $\omega$ is determined by inspecting the optimal regression fitting on the training datasets in our experiments.

\begin{figure}[tb]
    \centering
    \includegraphics[width=0.95\textwidth]{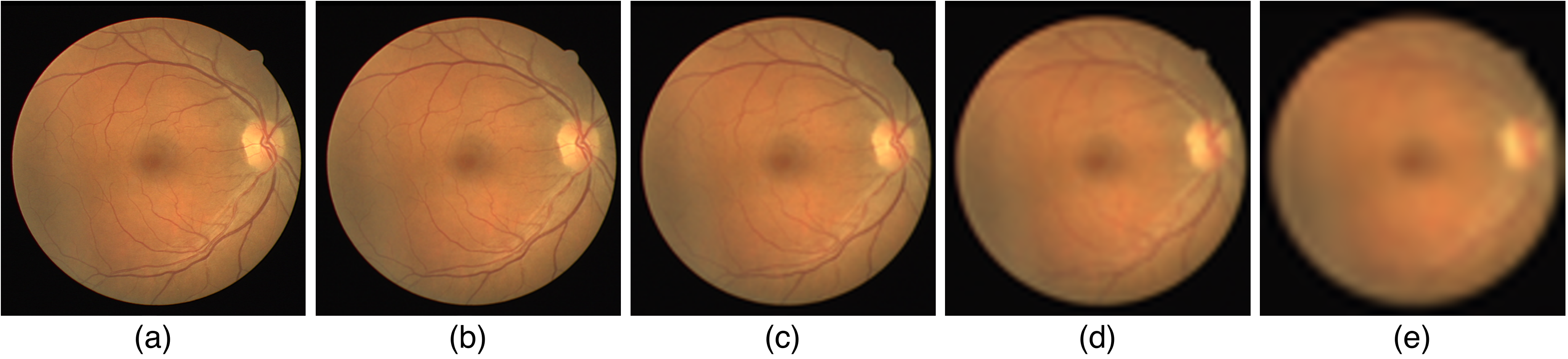}
    \caption{Information content degradation with image subsampling. In (a), an example image of the DRIVE dataset is shown. In (b), (c), (d), and (e), the original image is subsampled by a scale of 2, 4, 8, and 16, respectively. The subsampled image is upsampled to the original image scale to highlight the lost information content in the process of sampling.}
    \label{fig:lw-complex}
\end{figure}

\subsection{Layer-wise Complexity}
\label{ssec:lw-comp}
In CNNs, convolutional layers are stacked with intermediate sub-sampling operations in order to extract rich contextual features. Each sub-sampling operation reduces the input feature-map scale which is forwarded as input to the subsequent convolutional layers. Since every convolutional layer of a specific \textit{stage} of a CNN (in between two sub-sampling operations) extracts features from a specific feature-map scale, we explore complexity from an image scale perspective. Such a scale based complexity will enable us to understand the relative information content at a specific image scale which will be helpful in performing layer-wise pruning of a network (fine-grain pruning).  

In order to obtain layer-wise JPEG complexity, we extend the approach explained in Eq.~(\ref{eqn:jpeg}) by reformulating JPEG complexity as:

\begin{equation}
    J = \frac{size(Upsampled(Compressed(Subsampled(Image))))}{size(Image)}
\label{eqn:layer-wise}    
\end{equation}

Instead of using the ratio of the storage size of the compressed image and original image at a specific scale, we upsample every subsampled image before generating the complexity metric. Such an approach is used by following information theory to have a consistent frame of reference, where the input image to the network is considered as the base case with respect to which each calculation is performed. An example case of information content reduction with subsampling is shown in Fig.~\ref{fig:lw-complex}. Similar extension for layer-wise foreground density calculation is also performed.

\section{Network Parameter Calculation}
\label{sec:net-param}

The segmentation accuracy (e.g., F1 and IU scores) depends on many factors, such as the number of network weights, arrangement of network weights (network architecture), training methods, and certainly training data. From the discussions in Section~\ref{ssec:m_comp-select}, it is evident that accuracy is also closely related to the input dataset complexity. Keeping all the other variables (e.g., the network architecture and training method) unchanged, we can express the relationship between the segmentation accuracy and data complexity as $A = f(\theta, C)$, where $A$, $\theta$, and $C$ represent the segmentation accuracy, number of trainable weights, and training data complexity, respectively. For general networks, the function $f(\theta,C)$ can be rather complicate. But in general, segmentation accuracy is monotonically non-decreasing with respect to $\theta$ and $C$, i.e., $\frac{\partial f}{\partial \theta} \geq 0$ and $\frac{\partial f}{\partial C} \geq 0$. 

For CNNs, which are widely used for biomedical image segmentation, we observe (as discussed in Section~\ref{ssec:dod-cal}) that $\frac{\partial f}{\partial \log \theta}$ can be approximated by a linear function of $C$. That is, 
\begin{equation}
    \frac{\partial f}{\partial \log \theta} = \lambda C + \delta
    \label{eqn:main}
\end{equation}
for a constant $\lambda$ which reflects the {\it degree of degradation}. Given the linear dependency of $\frac{\partial f}{\partial \log \theta}$ on $C$, if $C$, $\lambda$, and $\log \theta$ are known, then it is straightforward to compute the change in accuracy or in the number of trainable network weights, when the other factors are provided. The value of $\lambda$ is network-dependent, and can be obtained by performing systematic network compression and tracking the corresponding change in accuracy.

\begin{figure}[tb]
    \centering
    \includegraphics[width=0.8\textwidth]{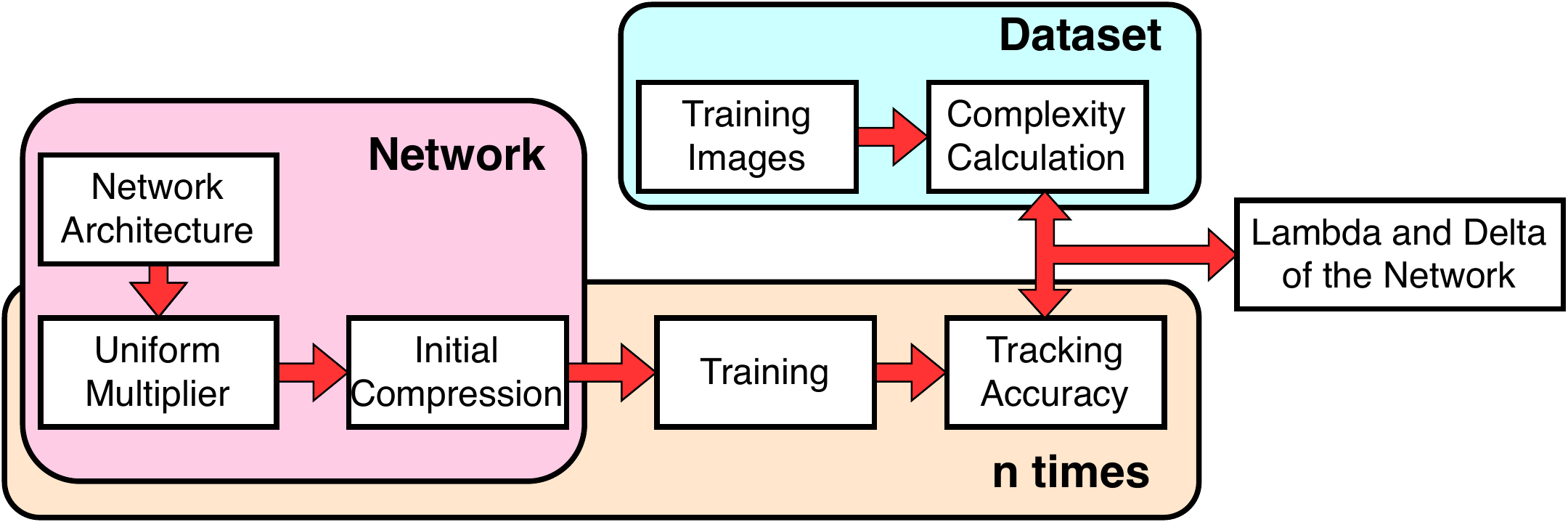}
    \caption{Degree of degradation calculation for a specific network architecture. By selecting $n$ different $\alpha$ values, we prune and track the accuracy drop of the architecture under consideration $n-times$ for a robust approximation. The drop in accuracy is then mapped to the complexity of the dataset for which it is trained to generate the value of $\lambda$ and $\delta$.}
    \label{fig:lambda-delta-calc}
\end{figure}

To obtain a compressed network, a widely used method is to reduce the number of channels in the feature maps. Since a channel multiplier based uniform reduction on the number of feature maps is quite simple and performs very well \cite{morphnet}, we use it for our network compression. For CNNs, the stored weights (determining the memory usage) are the weights of the filters for each convolutional layer, which, when ignoring biases, can be calculated as 
\begin{equation}
   \theta = FM_i \times F_i^X \times F_i^Y \times FM_{i+1}
\end{equation}
where $FM_i$ and $FM_{i+1}$ are the numbers of channels in the input and output feature maps, $F_i^X$ and $F_i^Y$ are the dimensions of the filter. With a multiplier $\alpha \in (0,1]$, the number of network weights is reduced to 
\begin{equation}
    \theta^* = \alpha FM_i \times F_i^X \times F_i^Y \times \alpha FM_{i+1} = \alpha^2 \theta
\end{equation}
Note that for a given $\alpha$, one can reduce the number of weights by $\approx \alpha^2$ \cite{suraj}. 

Our proposed framework for \textit{degree of degradation} calculation is shown in Fig.~\ref{fig:lambda-delta-calc}. Using a specific $\alpha$ value, a thinner network architecture is generated. Training is performed on this thinner architecture and output accuracy is reported. In order to generate robust $\lambda$ and $\delta$ values, we repeat this procedure for multiple times with different $\alpha$ values. For every network, this operation has to be performed once, as we have assumed $\lambda$ and $\delta$ to be specific for a fixed network architecture.

\section{Design Constraints Inclusion}
\label{sec:sec-user-const}
When producing compressed networks for biomedical image segmentation, we consider two practical design scenarios: (1) memory-constrained best possible accuracy, and (2) accuracy-guided least memory usage. Case 1 with memory-constrained best possible accuracy represents scenarios in many embedded devices where there is a memory budget. The budget can be provided either as main memory usage or as disk space storage and the objective is to design a network which can achieve maximum possible accuracy under the memory budget constraint. Case 2 with accuracy-guided least memory usage represents scenarios where multiple processes are sharing a single resource. In such a setup, some processes can be considered as auxiliaries to certain higher priority main processes where it is \textit{ok} to compromise the accuracy of such auxiliary processes as long as it does not fall below a certain threshold. The budget can be provided as the accuracy threshold and the objective is to achieve least memory usage in order to free up resources for the main processes.        

For each user constraint we explore two directions to compress the network: (a) using a uniform multiplier, and (b) using a nonuniform layer-wise multiplier.

\subsection{Memory Constrained Best Possible Accuracy}
\label{ssec:mem_const_best_poss_acc}
The memory budget can be provided either as disk space budget or as main memory budget. The disk space budget sets an upper bound on the number of total trainable weights that the compressed network can have. The main memory budget similarly sets an upper bound on the number of total trainable weights in the compressed network. However, in this case, besides considering the number of bits for each weight, one must also take into consideration the sizes of intermediate feature maps since they also occupy the main memory when performing convolution operation. We provide a detailed formulation for disk space budget constraint while only highlighting the modifications necessary for main-memory budget consideration.

\textbf{Uniform Multiplier:}
Given a disk space budget in MB, we first determine the number of trainable network weights, $\theta^*$, for the compressed network, based on the number of bits for each weight. Then a uniform multiplier $\alpha$ can be computed as 
\begin{equation}
    \alpha = \sqrt{\frac{\theta^*}{\theta}}
    \label{eqn:uni_mul}
\end{equation}
where $\theta$ is the number of network weights for the uncompressed original network model. Similarly for a main memory budget, $\theta^*$ can be calculated. However, taking intermediate feature-maps into consideration, the uniform multiplier in this case will be: 
\begin{equation}
    \alpha = \frac{\theta^*}{\theta}.
\end{equation}

\textbf{Nonuniform Multiplier:}
We want to formulate an approach for nonuniform layer-wise multiplier for effective pruning where each convolutional layer is pruned based on the layer-wise complexity of the image scale from which it extracts feature. To simplify notations, we consider a CNN with two convolutional layers, but similar results can be derived for any other CNN.   

Consider a CNN with two convolutional layers and a sub-sampling operation in between. The first convolutional layer is associated with image complexity (C1) while second convolutional layer has C2 as the associated image complexity. As segmentation accuracy is defined for the whole network, for a different degree of pruning of layer 1 ($\theta_1 \rightarrow \theta_1^*$) and layer 2 ($\theta_2 \rightarrow \theta_2^*$), we can rewrite Eq.~(\ref{eqn:main}) as  

\begin{equation}
     (\lambda C_1 + \delta)\Delta\log\theta_1 = (\lambda C_2 + \delta)\Delta\log\theta_2 
\end{equation}
 
\begin{equation}
    \implies (\lambda C_1 + \delta)(\log\theta_1 - \log\theta_1^*) = (\lambda C_2 + \delta)(\log\theta_2 - \log\theta_2^*)
\end{equation}

\begin{equation}
    \implies (\lambda C_1 + \delta)(\log\frac{\theta_1}{\theta_1^*}) = (\lambda C_2 + \delta)(\log\frac{\theta_2}{\theta_2^*})
    \label{eqn:theta-to-alpha}
\end{equation}

Using Eq.~(\ref{eqn:uni_mul}), Eq.~(\ref{eqn:theta-to-alpha}) can be rewritten as: 
\begin{equation}
     (\lambda C_1 + \delta)(\log\alpha_1) = (\lambda C_2 + \delta)(\log\alpha_2)
     \label{eqn:combn}
\end{equation}

Further, we can rewrite Eq.~(\ref{eqn:uni_mul}) for the two convolutional layer CNN as:
\begin{equation}
 \theta^* = \alpha^2 \theta = \alpha^2 \theta_1 + \alpha^2 \theta_2   
 \label{eqn:uni_mem}
\end{equation}
where $\theta_1$ and $\theta_2$ are the weights associated with the first and second convolutional layers, respectively. However, with nonuniform multipliers associated with each individual layer (i.e., $\alpha_1$ and $\alpha_2$ for the first and second layers, respectively), Eq.~(\ref{eqn:uni_mem}) results in 

\begin{equation}
 \theta^* = \alpha_1^2 \theta_1 + \alpha_2^2 \theta_2   
 \label{eqn:uni_mem_alpha}
\end{equation}

Using Eq.~(\ref{eqn:uni_mem_alpha}) and Eq.~(\ref{eqn:combn}), $\alpha_1$ and $\alpha_2$ can be determined when $\lambda$, $\delta$, $C_1$, $C_2$, $\theta_1$, $\theta_2$, and $\theta^*$ are known. For a main memory budget, a similar approach can be used as Eq.~(\ref{eqn:combn}) is unchanged for both the cases. The only modification is on Eq.~(\ref{eqn:uni_mem_alpha}) which becomes

\begin{equation}
     \theta^* = \alpha_1 \theta_1 + \alpha_2 \theta_2   
 \label{eqn:non_uni_mem_alpha}
\end{equation}

\subsection{Accuracy Guided Least Memory Usage}
\label{ssec:acc_gud_lst_mem_usg}
Provided the lowest acceptable accuracy ($A_{min}$) as a percentage of best possible accuracy ($A_{best}$), our objective is to generate a model with least memory usage. We consider both uniform and nonuniform layer-wise multipliers for this case and provide implementation details as follows.

\textbf{Uniform Multiplier:} For a given accuracy threshold, $\Delta A$ (= $A_{best} - A_{min}$) can be computed. Using complexity C, and network $\lambda$ and $\delta$, change in number of network trainable weights can be computed as 
\begin{equation}
    \Delta \log\theta = \frac{\Delta A}{\lambda C + \delta}
\end{equation}

\begin{equation}
    \log\theta - \log{\theta^*} = \frac{A_{best} - A_{min}}{\lambda C + \delta}
    \label{eqn:uni-acc}
\end{equation}

Using Eq.~(\ref{eqn:uni-acc}), $\theta^*$ can be calculated. The uniform multiplier $\alpha$, can be calculated using $\theta$, and $\theta^*$ values using Eq.~(\ref{eqn:uni_mul}) as discussed in Section~\ref{ssec:mem_const_best_poss_acc}.

\textbf{Layer-wise Multiplier:} 
For nonuniform layer-wise multiplier determination we formulate the problem using the two layer CNN as explained in Section~\ref{ssec:mem_const_best_poss_acc}. We divide the layer-wise multiplier determination task into two sub-problems each associating with one CNN layer. Each sub-problem represent a network extracting features from an image with associated complexity of $C_i$. Using $C_i$, $\lambda$, and $\delta$ (as the network structure is the same), we can determine $\theta_i^*$ as determined in Eq.~(\ref{eqn:uni-acc}), i.e., 

\begin{equation}
    \log\theta - \log{\theta_i^*} = \frac{A_{best} - A_{min}}{\lambda C_i + \delta}
    \label{eqn:non_uni-acc}
\end{equation}

Essentially, a system of equations are generated associating each convolutional layer with respective layer-wise complexity. Using $\theta_i^*$, and $\theta$ values, $\alpha_i$ can be calculated which is used to compress for $i^{th}$ layer specifically. Intuitively, layers dealing with images of higher complexities are compressed less, while layers extracting features from less complex images are compressed more.

\section{Experimental Evaluation}
\label{sec:exp}
We first provide the details of the datasets used in our experiments in Section~\ref{ssec:dataset}. Network architectures are described in Section~\ref{ssec:network}. The degree of degradation calculation is shown in Section~\ref{ssec:dod-cal}. Finally, user constraint based network compression is explained in Section~\ref{ssec:user-const}.

\begin{table*}[tb]
\caption{JPEG complexity calculation.}
\begin{center}
\scalebox{1}{
\begin{tabular}{ c | c | c | c  | c | c }
\hline
\rule{0pt}{8pt} Scale & Wing disk & DRIVE & Melanoma & Lymph node & CHASE\_DB1 \\ \hline
Input & 0.0279 & 0.0362 & 0.0642 & 0.1518 & 0.2826  \\ 
Input/$2^1$ & 0.0187 & 0.0303 & 0.0459 & 0.0857 & 0.2204 \\
Input/$2^2$ & 0.0175 & 0.0284 & 0.0361 & 0.0655 & 0.1971   \\
Input/$2^3$ & 0.0166 & 0.0269 & 0.0296 & 0.0496 & 0.1789  \\
Input/$2^4$ & 0.0156 & 0.0255 & 0.0250 & 0.0375 & 0.1636 \\
\hline
\end{tabular}}
\end{center}
\label{tab:j_complexity_values}
\end{table*}

\begin{table*}[tb]
\caption{{\revised JB complexity calculation for the lymph node dataset.}}
\begin{center}
\scalebox{1}{
\begin{tabular}{ c | c | c | c  | c | c }
\hline
\rule{0pt}{8pt} Scale & J & B & JB (U-Net) & JB (CUMedVision) & JB (UCU-Net) \\ \hline
Input & 0.1518 & 0.0812 & 0.1306 & 0.1359 & 0.1483  \\ 
Input/$2^1$ & 0.0857 & 0.0813 & 0.0844 & 0.0847 & 0.0855 \\
Input/$2^2$ & 0.0655 & 0.0811 & 0.0702 & 0.0690 & 0.0663   \\
Input/$2^3$ & 0.0496 & 0.0809 & 0.0590 & 0.0566 & 0.0512  \\
Input/$2^4$ & 0.0375 & 0.0799 & 0.0502 & 0.0470 & 0.0396 \\
\hline
\end{tabular}}
\end{center}
\label{tab:jb_complexity_values}
\end{table*}

\subsection{Datasets and Complexities}
\label{ssec:dataset}
We experiment with five biomedical image datasets of different modalities. In the \textbf{DRIVE} dataset~\cite{drive}, 40 fundus images are provided for retinal vessel segmentation. 20 images are used for training and the other 20 images are used for evaluation. In the \textbf{CHASE\_DB1} dataset~\cite{chasedb1}, 28 fundus images are provided for retinal vessel segmentation without any specific train-test split. Following~\cite{xia,mishra}, we use 20 images for training and the remaining 8 images for evaluation. \textbf{Melanoma} segmentation using the ISIC 2017 skin lesion dataset~\cite{isic} contains 2000 training, 150 validation, and 600 test RGB images for melanoma segmentation. Noticing the smaller validation set, we merge the training and validation sets and randomly select 20\% of the merged set for validation as in \cite{me_cvpr_data}. The \textbf{lymph node} dataset contains ultrasound images of the lymph node areas of 237 patients. Following~\cite{yizhe}, we use 137 images for training (20\% for validation) and the rest for testing, assuring no identity overlap. \textbf{Wing disc} pouches of fruit flies are used to study organ development~\cite{peixian,suraj}. 996 grayscale wing disc pouch images are investigated by using 889 images for training (20\% for validation) and 107 images for testing.

In Table~\ref{tab:j_complexity_values}, JPEG complexity values calculated for these five biomedical image datasets are shown. CHASE\_DB1 has the highest JPEG complexity among all the datasets while the wing disk dataset is considered as the least complex dataset for our experiments. Further, complexity values for different image scales are also provided. Observe that with subsampling operations, JPEG complexity decreases, indicating reduction in information content. 

{\revised As discussed in Section~\ref{ssec:m_comp-select}, for JB calculation, $\omega$ is determined by examining the optimal regression fitting between $\partial A / \partial \log\theta$ vs JB, where the accuracy $A$ = IU. The $\omega$ value resulting in the best regression fitting (the best $R^{2}$) is used for JB calculation. For U-Net, CUMedVision, and UCU-Net, the $\omega$ values thus found are 0.7, 0.775, and 0.95, respectively. In Table~\ref{tab:jb_complexity_values}, the JB values for the lymph node dataset are shown. Observe that with scaling, the blob density remains relatively constant.}

\begin{figure}[tb]
    \centering
    \includegraphics[width=0.7\textwidth]{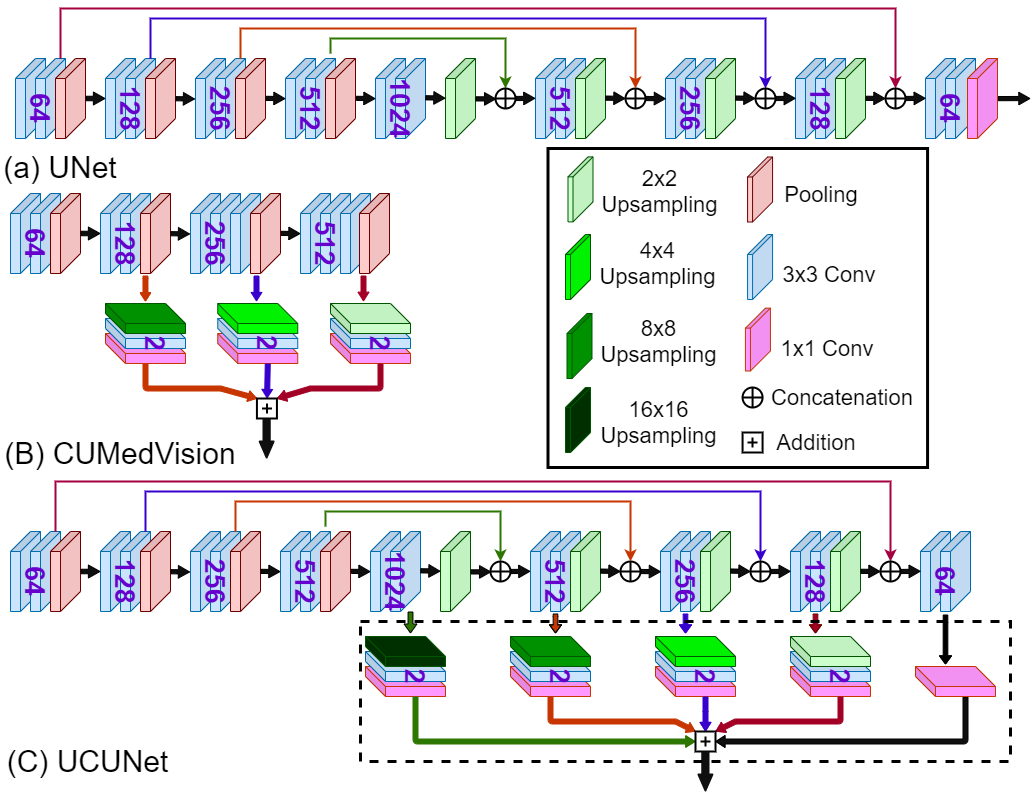}
    \caption{CNN architectures used in our experiments.}
    \label{fig:network-arch}
\end{figure}

\subsection{Network Architecture and Setup}
\label{ssec:network}
Three common networks (shown in Fig.~\ref{fig:network-arch}), with an encoder-decoder architecture, for biomedical image segmentation are used in our experiments. U-Net~\cite{unet} and CUMedVision~\cite{cumednet} are shown in Fig.~\ref{fig:network-arch}(a) and Fig.~\ref{fig:network-arch}(b), respectively. In Fig.~\ref{fig:network-arch}(c), UCU-Net~\cite{mishra} architecture is highlighted which has a similar encoder as U-Net. For the decoder, UCU-Net combines the U-Net and CUMedVision decoders to generate an architecture with superior contextual information flow \cite{retinanet, fpn, uncert}.

The experiments utilize the PyTorch framework with the \textit{He} initialization~\cite{he_init}. To limit overfitting on a small training set, data augmentation is performed using random flipping and rotation. The training uses the Adam~\cite{adam} optimizer ($\beta_1 = 0.9, \beta_2 = 0.999, \epsilon = 1\mathrm{e}{-10}$) with a fixed learning rate of 0.00002 using a cross-entropy based loss function. Experiments are performed on NVIDIA-TITAN and Tesla P100 GPUs for a number of epochs (CHASE\_DB1: 5000, DRIVE: 5000, Melanoma: 3000, lymph node: 5000, wing disk: 3000). The images are resized (CHASE\_DB1: 976 $\times$ 976~\cite{mishra}, DRIVE: 512 $\times$ 512~\cite{mishra}, Melanoma: 320 $\times$ 320~\cite{mel_320}, lymph node: 224 $\times$ 224, wing disc: 320 $\times$ 320), and the training uses 128$\times$128 size patches. The batch size for each case is selected as the maximum size permissible by the GPU. 

\begin{figure}[tb]
    \centering
    \includegraphics[width=0.8\textwidth]{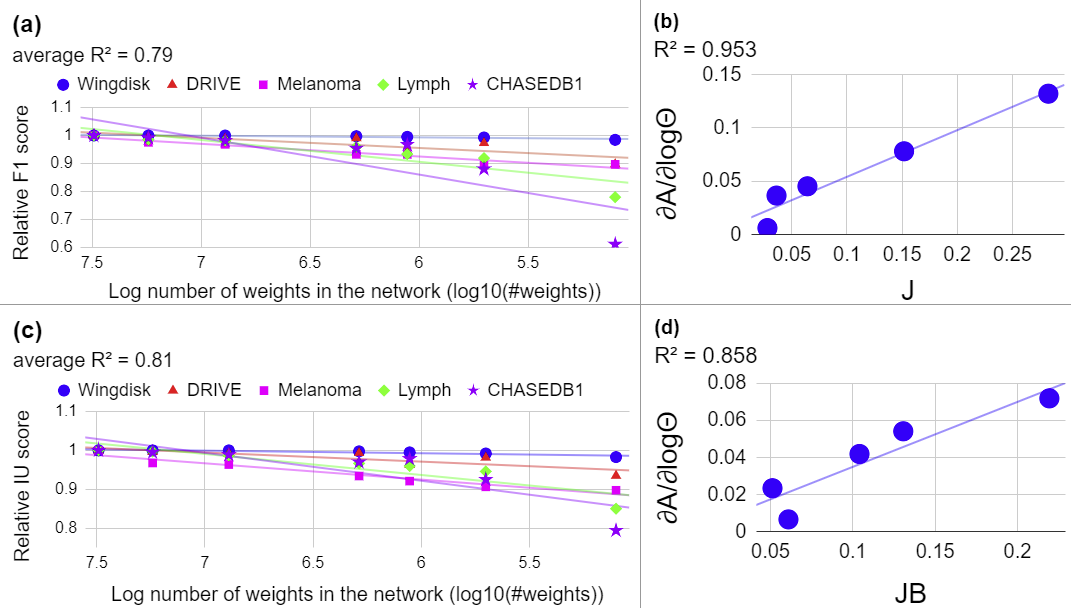}
    \caption{Degree of degradation ($\lambda$ and $\delta$) calculation for the U-Net architecture.}
    \label{fig:unet-plots}
\end{figure}

\begin{figure*}[tb]
    \centering
    \includegraphics[width=0.8\textwidth]{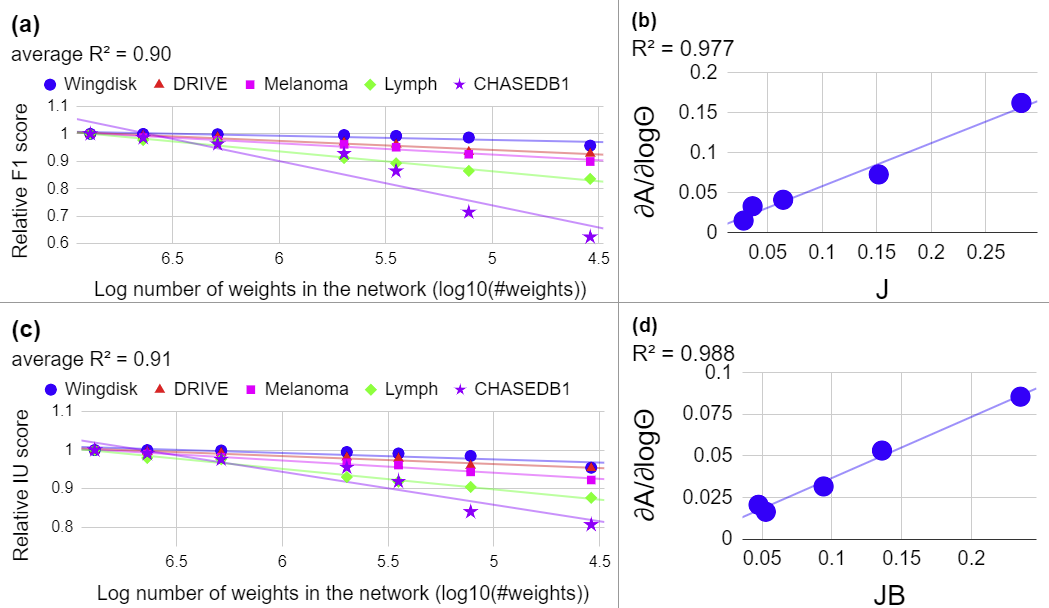}
    \caption{Degree of degradation ($\lambda$ and $\delta$) calculation for the CUMedVision architecture.}
    \label{fig:cunet-plots}
\end{figure*}

\begin{figure*}[tb]
    \centering
    \includegraphics[width=0.8\textwidth]{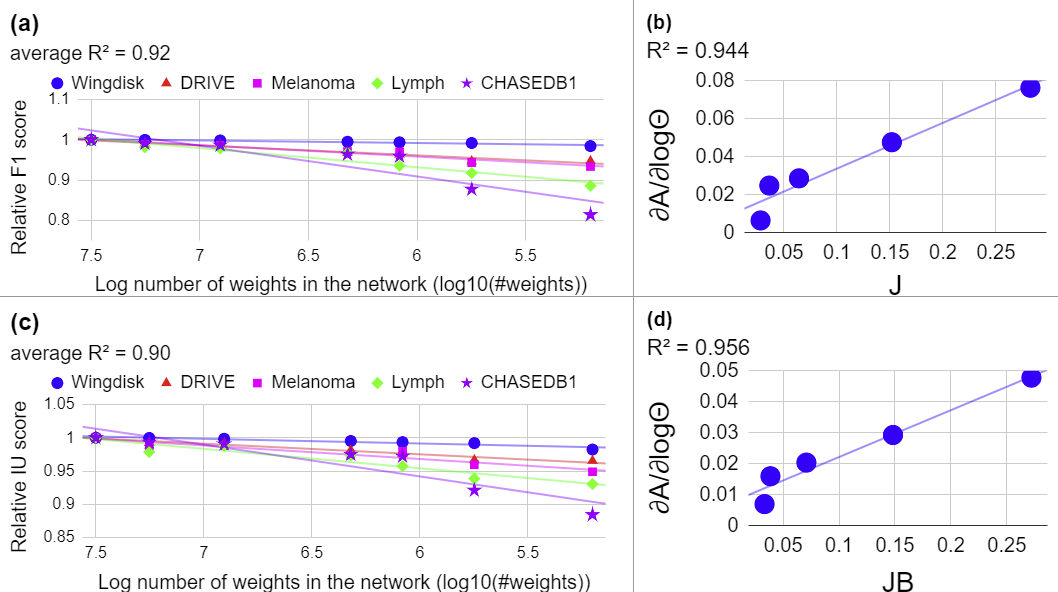}
    \caption{Degree of degradation ($\lambda$ and $\delta$) calculation for the UCU-Net architecture.}
    \label{fig:ucu-plots}
\end{figure*}

\begin{table}[tb]
\caption{$\lambda$ and $\delta$ calculation.}
\label{tab:lambda-delta}
\begin{center}
\begin{tabular}{ c | c  c | c  c }
\hline
\multicolumn{1}{c}{} & \multicolumn{2}{|c}{F1} & \multicolumn{2}{|c}{IU} \\
\hline
\rule{0pt}{6pt}Architecture & $\lambda$ & $\delta$ & $\lambda$ & $\delta$ \\ \hline
U-Net & 0.437 & 0.01030 & 0.349 & 0.000072 \\
CU-Net & 0.538 & 0.00427 & 0.366 & 0.000096 \\
UCU-Net & 0.241 & 0.00956 & 0.151 & 0.007080 \\
\hline
\end{tabular}
\end{center}
\end{table}

\subsection{Degree of Degradation Calculation}
\label{ssec:dod-cal}
As explained in Section~\ref{sec:net-param}, for the degree of degradation calculation, we systematically compress a given network architecture and track the accuracy degradation caused by the network compression. Then we map the dataset complexity ($C$) with the accuracy degradation caused by compression ($\frac{\partial A}{\partial \log \theta}$) to determine the degree of degradation (i.e., $\lambda$ and $\delta$).

For simpler calculations maintaining the integer filter (channel) values, $\alpha \in \{1, 0.75, 0.5, 0.25, 0.1875,\\ 0.125, 0.0625\}$ are used for network compression. The compressed network generated after multiplying $\alpha$ uniformly across all the convolutional layers is trained and the corresponding accuracy is reported. In Fig.~\ref{fig:unet-plots}(a), drop in F1-score is plotted against the log number of trainable weights of the U-Net architecture. Each data point corresponds to the relative F1 score (i.e., $\frac{F1_{\alpha}}{F1_{\alpha=1}}$) for a specific network weight (i.e., a specific $\alpha$). We repeat this procedure for all of the five datasets to generate the trend. The slope of the linear trend line best fitting the set of data points for a single dataset (i.e., slope = $\frac{\partial A}{\partial \log \theta}$) is calculated. In Fig.~\ref{fig:unet-plots}(b), the calculated slope for each dataset is plotted against the complexity of that specific dataset. The straight line best fitting the distributions of the points essentially represents the degree of degradation for that specific network architecture as the equation of the regressed line is $\frac{\partial A}{\partial \log \theta} = \lambda C + \delta$, where $\lambda$ and $\delta$ are the slope and the y-intercept of the regressed trend line. {\revised Similar calculations for the IU score degradation determination are shown in Fig.~\ref{fig:unet-plots}(c) and Fig.~\ref{fig:unet-plots}(d) (i.e., the drop in IU score is shown in Fig.~\ref{fig:unet-plots}(c) and degree of degradation for IU is shown in Fig.~\ref{fig:unet-plots}(d))}. Experiments on CUMedVision and UCU-Net are shown in Fig.~\ref{fig:cunet-plots} and Fig.~\ref{fig:ucu-plots}, respectively. For all the three examined networks, the calculated $\lambda$ and $\delta$ values {\revised associated with the F1 and IU scores} are tabulated in Table~\ref{tab:lambda-delta}.

\begin{table*}[tb]
\caption{Segmentation results for the CHASE\_DB1 and DRIVE datasets for different $\alpha$ values.}
\begin{center}
\scalebox{1}{
\begin{tabular}{ c | c  c  c  c  c  c | c  c  c  c  c  c}
\hline
\multicolumn{1}{c|}{} & \multicolumn{6}{c|}{CHASE\_DB1~\cite{chasedb1}} & \multicolumn{6}{c}{DRIVE~\cite{drive}} \\
\hline
\rule{0pt}{8pt} Method & AUC & Acc & Spe & Sen & F1 & IU & AUC & Acc & Spe & Sen & F1 & IU \\ \hline
$\alpha$ = 1 & 0.9797 & 0.9740 & 0.9915 & 0.7121 & 0.7748 & 0.8029 & 0.9762 & 0.9673 & 0.9903 & 0.7308 & 0.7940 & 0.8128 \\ 
$\alpha$ = 0.75 & 0.9795 & 0.9736 & 0.9916 & 0.7041 & 0.7684 & 0.7989 & 0.9763 & 0.9668 & 0.9902 & 0.7265 & 0.7901 & 0.8102 \\  
$\alpha$ = 0.5 & 0.9801 & 0.9724 & 0.9896 & 0.7151 & 0.7602 & 0.7939 & 0.9782 & 0.9664 & 0.9910 & 0.7133 & 0.7857 & 0.8067 \\
$\alpha$ = 0.25 & 0.9770 & 0.9710 & 0.9921 & 0.6576 & 0.7395 & 0.7787 & 0.9796 & 0.9662 & 0.9896 & 0.7257 & 0.7864 & 0.8073 \\
$\alpha$ = 0.1875 & 0.9794 & 0.9717 & 0.9915 & 0.6777 & 0.7498 & 0.7855 &  0.9748 & 0.9629 & 0.9930 & 0.6522 & 0.7494 & 0.7826 \\
$\alpha$ = 0.125 & 0.9737 & 0.9667 & 0.9930	& 0.5749 & 0.6823 & 0.7427 & 0.9786 & 0.9648 & 0.9908 & 0.6978 & 0.7727 & 0.7977 \\
$\alpha$ = 0.0625 & 0.8866 & 0.9550 & 0.9953 & 0.3472 & 0.4739 & 0.6376 & 0.9684 & 0.9590 & 0.9932 & 0.6068 & 0.7125 & 0.7597 \\
\hline
\end{tabular}}
\end{center}
\label{tab:chasedb1_drive}
\end{table*}

\begin{figure*}[tb]
  \centering
  \centerline{\includegraphics[width=0.98\textwidth]{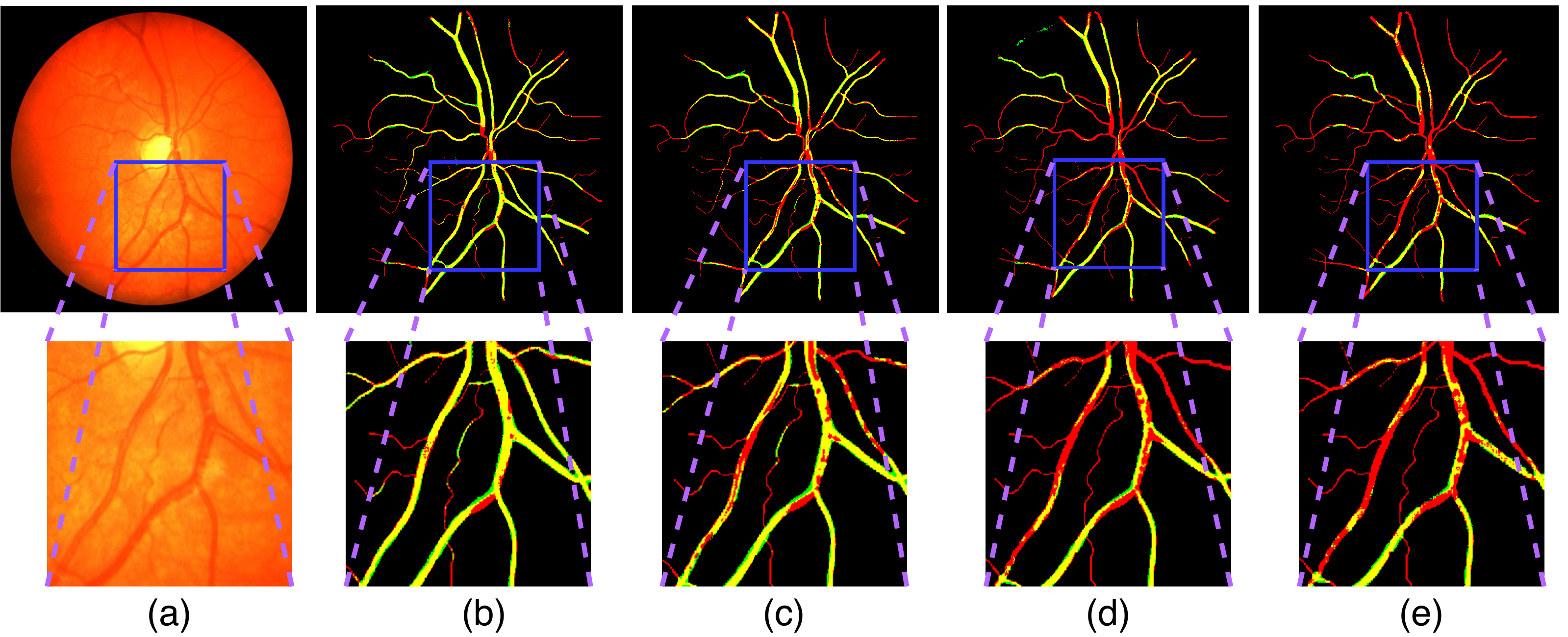}}
\caption{(a) An example CHASE\_DB1 image. In (b), (c), (d), and (e), ground truth and segmentation outputs are shown in Red and Green channel, respectively (Yellow = Red + Green). Each segmentation output is generated using a different $\alpha$ value, i.e., for (b) $\alpha = 1$, for (c) $\alpha = 0.5$, for (d) $\alpha = 0.1875$, and for (e) $\alpha = 0.0625$. Observe the degradation in the foreground segmentation with reducing $\alpha$ value.}
\label{fig:chasedb1_comp}
\end{figure*}

\begin{figure*}[tb]
  \centering
  \centerline{\includegraphics[width=0.98\textwidth]{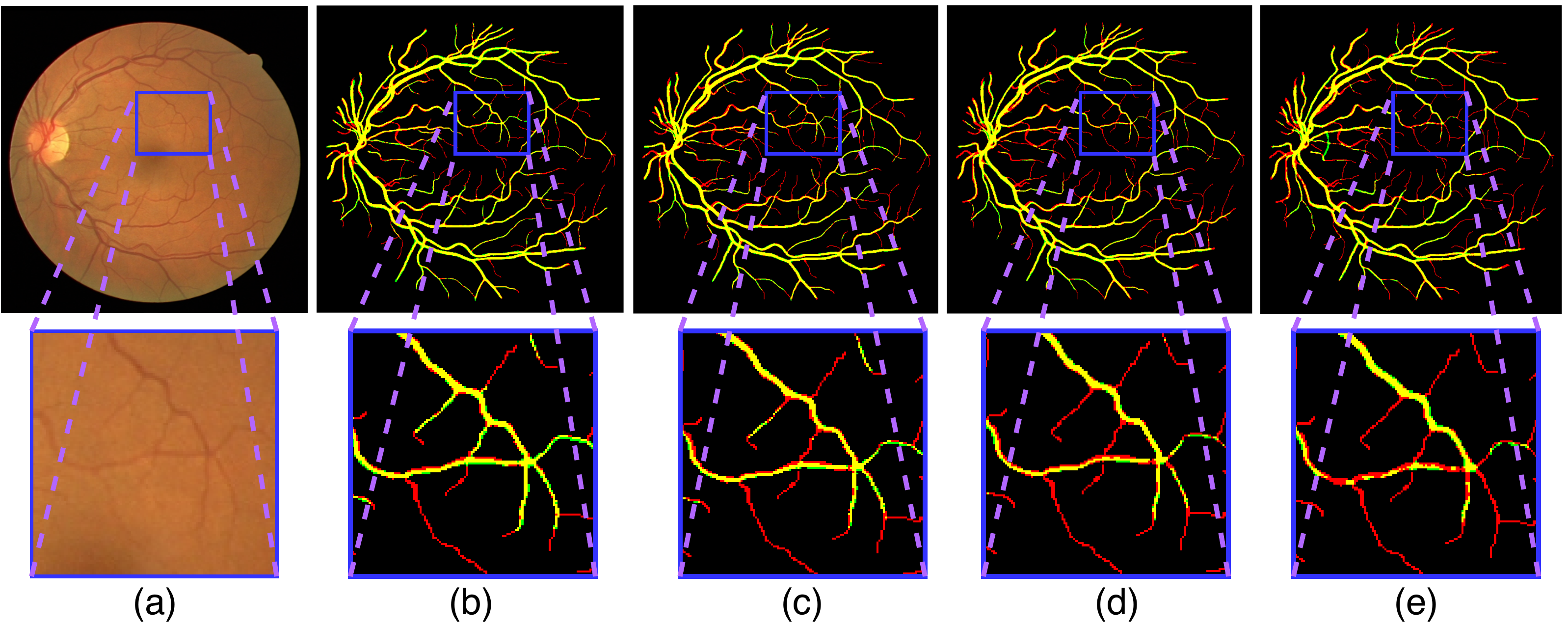}}
\caption{(a) An example DRIVE image. In (b), (c), (d), and (e), ground truth and segmentation outputs are shown in Red and Green channel, respectively (Yellow = Red + Green). Each segmentation output is generated using a different $\alpha$ value, i.e., for (b) $\alpha = 1$, for (c) $\alpha = 0.5$, for (d) $\alpha = 0.1875$, and for (e) $\alpha = 0.0625$. Observe the degradation in the foreground segmentation with reducing $\alpha$ value.}
\label{fig:drive_comp}
\end{figure*}

Accuracies obtained on the CHASE\_DB1 and DRIVE datasets for different $\alpha$ settings on the U-Net architecture are highlighted in Table~\ref{tab:chasedb1_drive}. Observe that for different $\alpha$ settings, Specificity (Spe) and Accuracy (Acc) do not show any change. Such behavior can be attributed to the highly imbalance nature of these datasets. Higher number of background pixels dominate the smaller foreground pixels and are not reflected significantly as accuracy drop (Acc). However, with network compression Sensitivity (Sen) decreases significantly and is also shown in Fig.~\ref{fig:chasedb1_comp} and Fig.~\ref{fig:drive_comp}. The background segmentation quality does not degrade significantly resulting in higher specificity while the foreground segmentation quality shows significant degradation resulting in poor sensitivity for these two datasets. However, since both F1 and IU metrics takes both background and foreground into consideration (as explained in Section~\ref{sec:net-param}), the drop in accuracy with compression is correctly captured by these two metrics.

\begin{figure*}[tb]
  \centering
  \centerline{\includegraphics[width=0.9\textwidth]{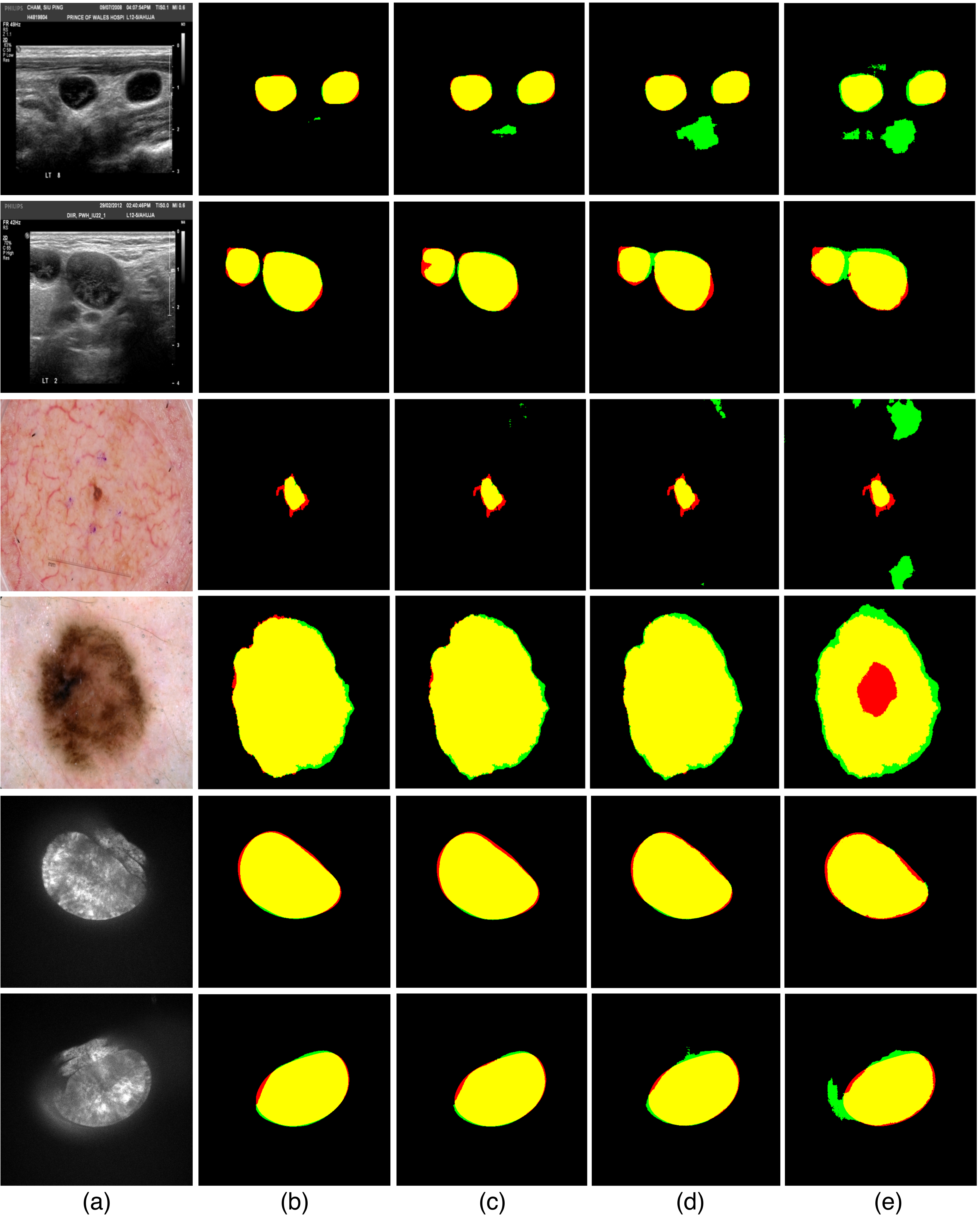}}
\caption{Qualitative results. In the top two rows, lymph node images are shown. In the middle two rows and the bottom two rows, melanoma, and wing disk images are shown, respectively. (a) An example image. In (b), (c), (d), and (e), ground truth and segmentation outputs are shown in Red and Green channel, respectively (Yellow = Red + Green). Each segmentation output is generated using a different $\alpha$ value, i.e., for (b) $\alpha = 1$, for (c) $\alpha = 0.5$, for (d) $\alpha = 0.1875$, and for (e) $\alpha = 0.0625$. Observe the degradation in the foreground segmentation with reducing $\alpha$ value.}
\label{fig:comb_comp}
\end{figure*}

Accuracy obtained for melanoma dataset for different $\alpha$ settings on U-Net architecture are highlighted in Table~\ref{tab:melanoma_lymph_wing}. Similar results are also obtained for lymph node and wing disc datasets. Example cases showing qualitative results for different $\alpha$ settings are shown in Fig.~\ref{fig:comb_comp}.

\begin{table*}[tb]
\caption{Segmentation results for melanoma dataset for different $\alpha$ values.}
\begin{center}
\scalebox{1}{
\begin{tabular}{ c | c  c  c  c  c  c }
\hline
\multicolumn{1}{c|}{} & \multicolumn{6}{c}{Melanoma} \\
\hline
\rule{0pt}{8pt} Method & Jac & Acc & Spe & Sen & F1 & IU \\ \hline
$\alpha$ = 1 & 0.7648 & 0.9252 & 0.9778 & 0.8193 & 0.8444 & 0.8362 \\ 
$\alpha$ = 0.75 & 0.7226 & 0.9141 & 0.9693 & 0.7844 & 0.8238 & 0.8090\\
$\alpha$ = 0.5 & 0.7176 & 0.9107 & 0.9751 & 0.7683 & 0.8177 & 0.8058 \\
$\alpha$ = 0.25 & 0.6854 & 0.8990 & 0.9543 & 0.7722 & 0.7875 & 0.7809 \\
$\alpha$ = 0.1875 & 0.6604 & 0.8956 & 0.9780 & 0.7151 & 0.7874 & 0.7700 \\
$\alpha$ = 0.125 & 0.6506 & 0.8888 & 0.9442 & 0.7588 & 0.7548 & 0.7576 \\
$\alpha$ = 0.0625 & 0.6324 & 0.8860 & 0.9731 & 0.6984 & 0.7574 & 0.7504 \\
\hline
\end{tabular}}
\end{center}
\label{tab:melanoma_lymph_wing}
\end{table*}

\subsection{Design Constraints Consideration}
\label{ssec:user-const}
Using the design constraints as explained in Section~\ref{sec:sec-user-const}, we formulate two test cases to evaluate the effectiveness of our proposed framework.

\textbf{Test case 1 (memory-constrained best possible accuracy).} We consider a disk space budget of 1 MB for a U-Net architecture on the lymph node dataset. Our objective is to obtain a U-Net type architecture which can achieve the best possible accuracy for the disk-space budget of 1MB. Following the framework provided in Section~\ref{ssec:mem_const_best_poss_acc} we examine both the uniform and layer-wise multipliers for the experiments. The uniform multiplier ($\alpha$) is determined using Eq.~(\ref{eqn:uni_mul}) while the layer-wise multipliers are determined using Eq.~(\ref{eqn:combn}) and Eq.~(\ref{eqn:uni_mem_alpha}) (modified for the U-Net architecture). Complexity (C), $\lambda$ and $\delta$ values are used as shown in Table~\ref{tab:j_complexity_values} and Table~\ref{tab:lambda-delta}. The layer-wise U-Net filter arrangements for both the uniform and layer-wise multiplier cases are shown in Table~\ref{tab:test-case-filters}. Results of both the cases are given in Table~\ref{tab:test-case1-acc}. For the same disk space budget, layer-wise multiplier achieves a better F1 score than the uniform multiplier based approach. {\revised Similar to F1 score, layer-wise multiplier achieves a better IU score compared to the uniform multiplier based approach. Observe that compared to F1, IU shows relatively lower degradation with compression. This can be attributed to the degree of degradation associated with both the accuracy metrics. As shown in Table~\ref{tab:lambda-delta}, lambda (slope) associated with F1 is higher compared to that with the IU score.}

\begin{table*}[tb]
\caption{Layer-wise U-Net encoder filter arrangement for test-cases.}
\begin{center}
\scalebox{1}{
\begin{tabular}{ c | c | c  c  | c  c }
\hline
\multicolumn{1}{c|}{} & \multicolumn{1}{c|}{} & \multicolumn{2}{c|}{Test case 1} & \multicolumn{2}{c}{Test case 2} \\
\hline
\rule{0pt}{8pt} Conv-layer & U-Net & Uniform & Layer-wise & Uniform & Layer-wise \\ \hline
Layer 1 & 64 & 4 & 20 & 30 & 30  \\ 
Layer 2 & 64 & 4 & 20 & 30 & 30 \\
Layer 3 & 128 & 8 & 19 & 60 & 39   \\
Layer 4 & 128 & 8 & 19 & 60 & 39  \\
Layer 5 & 256 & 16 & 25 & 120 & 59 \\
Layer 6 & 256 & 16 & 25 & 120 & 59  \\
Layer 7 & 512 & 32 & 29 & 240 & 85  \\
Layer 8 & 512 & 32 & 29 & 240 & 85  \\
Layer 9 & 1024 & 65 & 33 & 480 & 119 \\
Layer 10 & 1024 & 65 & 33 & 480 & 119  \\
\hline
\end{tabular}}
\end{center}
\label{tab:test-case-filters}
\end{table*}

\begin{table*}[tb]
\caption{{\revised Segmentation accuracy of test case 1 with a constraint of $\log\theta \leq 5.097$.}}
\begin{center}
\scalebox{1}{
\begin{tabular}{ c | c | c  c  c }
\hline
\rule{0pt}{8pt} & Method & F1 & IU & $\log \theta$ \\ \hline
& Base & 0.8644 & 0.8777 & 7.492 \\ 
\hline
\multirow{2}{*}{\centering Test case 1: Objective -- higher F1} & Uniform multiplier & 0.7739 & 0.8157 & 5.089 \\
& Layer-wise multiplier & 0.8165 & 0.8456 & 5.090\\
\hline
\end{tabular}}
\end{center}
\label{tab:test-case1-acc}
\end{table*}

\textbf{Test case 2 (accuracy-guided least memory usage).} We consider an example constraint of $F1_{compressed} \geq 95\%F1_{base}$ for a U-Net architecture on the lymph node dataset. Our objective is to obtain a U-Net architecture with the least disk-space usage while not dropping its accuracy below 95\%. Following the framework provided in Section~\ref{ssec:acc_gud_lst_mem_usg}, we examine both the uniform and layer-wise multipliers for experiments. The uniform multiplier is determined using Eq.~(\ref{eqn:uni-acc}) while the layer-wise multipliers are determined using Eq.~(\ref{eqn:non_uni-acc}) (modified for the specific architecture). Complexity (C), $\lambda$ and $\delta$ values are used as shown in Table~\ref{tab:j_complexity_values} and Table~\ref{tab:lambda-delta}. The layer-wise filter arrangements for both the uniform and layer-wise multiplier cases are shown in Table~\ref{tab:test-case-filters}. Results of both the cases are shown in Table~\ref{tab:test-case2-acc}. For the same accuracy threshold, layer-wise multiplier significantly outperforms uniform multiplier by compressing the network more. {\revised Similar to test case 1, relatively lower degradation in IU score can be attributed to its lower lambda value.}

\begin{table*}[tb]
\caption{{\revised Segmentation accuracy of test case 2 with a constraint of $F1_{min} \geq 0.8212$.}}
\begin{center}
\scalebox{1}{
\begin{tabular}{ c | c | c  c  c}
\hline
\rule{0pt}{8pt} & Method & F1 & IU & $\log \theta$ \\ \hline
& Base & 0.8644 & 0.8777 & 7.492  \\ 
\hline
\multirow{2}{*}{\centering Test case 2: Objective -- lower $\theta$} & Uniform multiplier & 0.8278 & 0.8641 & 6.834 \\
& Layer-wise multiplier & 0.8273 & 0.8424  & 5.925\\
\hline
\end{tabular}}
\end{center}
\label{tab:test-case2-acc}
\end{table*}

\begin{figure}[tb]
  \centering
 \includegraphics[width=0.8\textwidth]{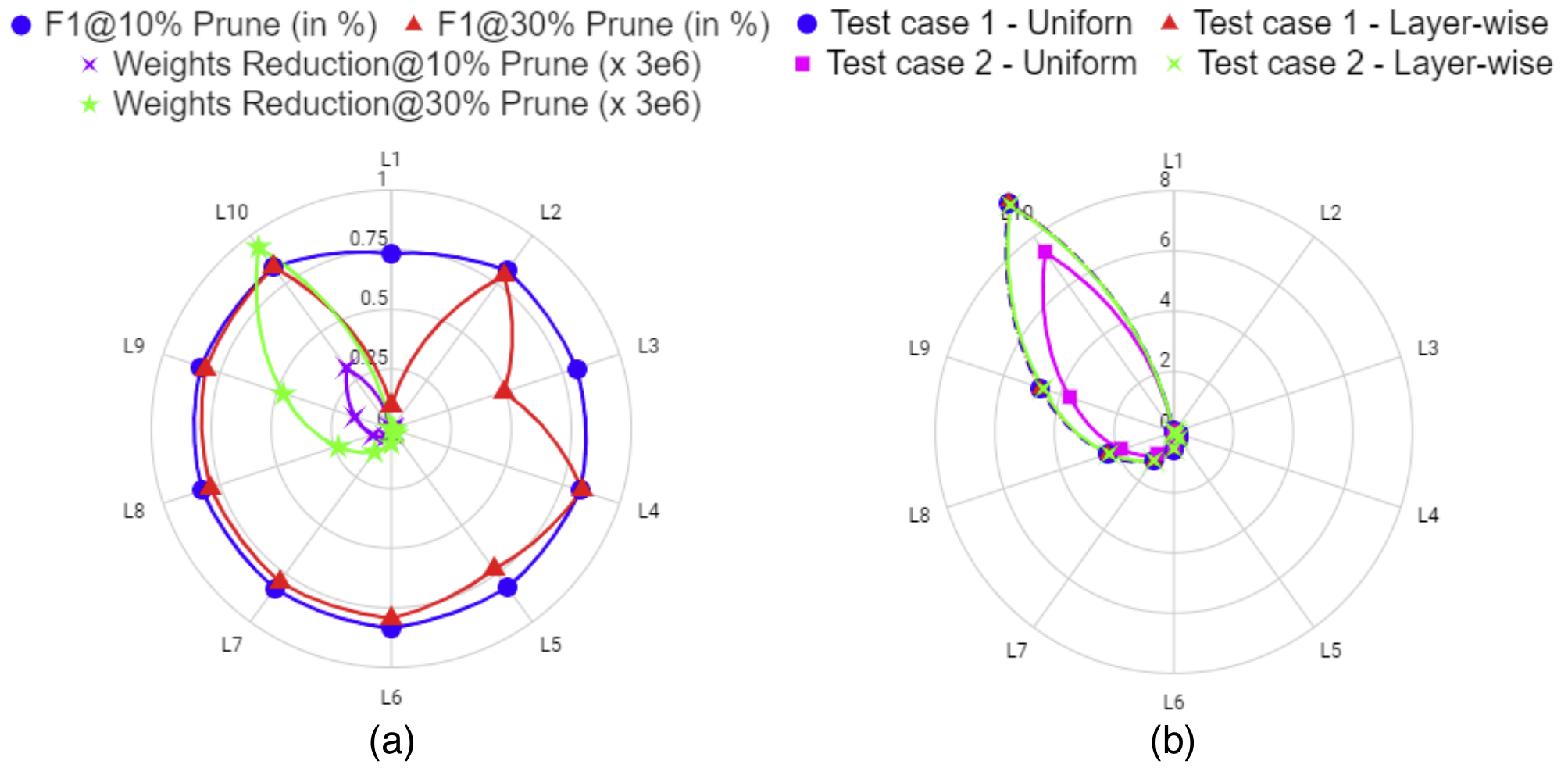}
\caption{(a) Random pruning of convolutional layers in the encoder (L1 to L10, weight reduction (in 3e6)). Pruning initial layers (e.g., L1) causes higher accuracy degradation for fewer pruned weights. Deeper layers (e.g., L10) can be pruned more with little negative effect on F1 score. (b) Pruning caused by our proposed method for both the test cases (weight reduction (in 1e6)).}
\label{fig:rand_prune_all}
\end{figure}

\section{Discussion}
\label{sec:discuss}
The degree of degradation for F1 score (as shown in Table~\ref{tab:lambda-delta}) reveals that smaller networks can be pruned less compared to larger networks as the accuracy degrades quickly for smaller networks (e.g., for CUMedVision). However, it is interesting to note that, for U-Net, the IU score degrades relatively similarly as the smaller CUMedVision architecture. We believe that this is caused by the decoder structure of U-Net, in which scale-wise information is not fused to generate the output. This implies that a scale-wise decoder (as in CUMedVision) is more efficient compared to the decoder arrangement of U-Net.

We perform random pruning of trainable weights for each layer of U-Net trained on the lymph node dataset. Results obtained are shown in Fig.~\ref{fig:rand_prune_all}(a). Pruning 30\% trainable weights in the initial layers of the network (e.g., L1), causes significant accuracy reduction. In comparison, pruning 30\% of deeper layer weights (which is significantly larger in count compared to the number of pruned weights for initial layers) are more robust as they do not adversely affect accuracy. In Fig.~\ref{fig:rand_prune_all}~(b), weight reduction achieved by our proposed framework for both the test cases are highlighted. Deeper layers are also more penalized by our approach, which intuitively verifies why initial layers of the network are more vital compared to deeper layers of the network.   

Additional experiments to verify the efficacy of our proposed framework are provided as follows. 

\begin{table*}[tb]
\caption{Prediction accuracy.}
\begin{center}
\scalebox{1}{
\begin{tabular}{ c | c | c  c  c  c  c}
\hline
\rule{0pt}{8pt} & Method & Predicted F1 & Achieved F1\\ \hline
\multirow{2}{*}{\centering Test case 2 (Ours)} & Uniform multiplier & $\geq$ 0.8212 & 0.8278 \\
& Layer-wise multiplier & $\geq$ 0.8212 & 0.8273\\
\hline
\multirow{2}{*}{\centering Test case 2 (Ours -- $\epsilon$)} & Uniform multiplier &  $<$ 0.8212 & 0.8201 \\
& Layer-wise multiplier & $<$ 0.8212 & 0.8077\\
\hline
\multirow{2}{*}{\centering Test case 2 (Ours + \cite{squeezenet})} & Uniform multiplier &  $<$ 0.8212 & 0.8073 \\
& Layer-wise multiplier & $<$ 0.8212 & 0.8029\\
\hline
\multirow{2}{*}{\centering Test case 2 (Ours + \cite{qbert})} & Uniform multiplier &  $<$ 0.8212 & 0.8194 \\
& Layer-wise multiplier & $<$ 0.8212 & 0.8107\\
\hline
\end{tabular}}
\end{center}
\label{tab:pred-acc}
\end{table*}

\subsection{Prediction Accuracy}
In Table~\ref{tab:pred-acc}, predicted F1 scores and achieved F1 scores for test case 2 are highlighted. For both the uniform and layer-wise multiplier based compressions, the conformity between the achieved F1 scores and the predicted F1 scores highlights the efficacy of our framework. Observe that uniform layer based compression achieves higher F1 scores compared to layer-wise multiplier compression. This is expected since the layer-wise multiplier based compression reduces more trainable weights by pruning a higher number of filters (which is the objective) while adhering to the accuracy constraint.   

To further verify the precision of our scheme, we intentionally reduce the multiplier value by a small amount (denoted by $\epsilon$ in Table~\ref{tab:pred-acc}). Such a reduction in the multiplier prunes one or two additional convolutional filters from the predicted amount in each convolutional layer. We observe from Table~\ref{tab:pred-acc} that, by further reducing the multiplier value, the network is unable to obey the accuracy constraint, validating the precision of our approach. For the layer-wise multiplier case, the accuracy drop is larger compared to the uniform multiplier case. We believe that there are fewer redundant/ineffective convolutional filters (also less trainable weights) in the network for the layer-wise multiplier case, and hence the drop is larger. Similar behavior is observed when further pruning by \cite{squeezenet} or performing \cite{qbert}.

\subsection{Comparison}
We perform Squeeze-Net~\cite{squeezenet} type compression on the U-Net architecture. Our experiments show that such an arrangement degrades accuracy. We think that the method in \cite{squeezenet} may not be very suitable for biomedical image segmentation as robust dense features are not extracted well by \textit{squeeze}-type architecture. Using \cite{nvidiapruning}, we randomly prune a few filters and fine-tune the network. After some iterations of pruning and fine-tuning, accuracy degrades significantly, as shown in Table~\ref{tab:comparison}. Results obtained using dynamic quantization~\cite{qbert} is also highlighted in Table~\ref{tab:comparison}. {\revised Experiments are also performed on U-Net~\cite{unet} using a uniform multiplier $\alpha$ inspired from~\cite{suraj}. With an $\alpha = 0.5$, the method in \cite{suraj} achieves an F1 score = 0.8561. However, additional trainable weights need to be used by \cite{suraj} compared to our proposed method. In \cite{pavlo}, channel pruning was explored by estimating the contribution of a filter to the final loss and iteratively pruning filters with smaller scores. Experiments with the best U-Net~\cite{unet} model generate an F1 score = 0.8383 using \cite{pavlo}. Neuron merging was explored in \cite{neuron_merging} to compensate for the information loss caused by filter pruning. As shown in Table~\ref{tab:comparison}, neuron merging is capable of improving the segmentation accuracy of the compressed networks generated using our proposed method. Compared to other techniques, our proposed framework generates a better compressed network while adhering to the accuracy constraint of test case 2. Using a pruning factor of 10\% with the $L_1$-norm as the pruning criterion, neuron merging used on the compressed model generated by our proposed layer-wise multiplier attains a better F1 score of 0.8354.}     

\begin{table*}[tb]
\caption{{\revised Comparison with several other compression schemes.}}
\begin{center}
\scalebox{1}{
\begin{tabular}{ c | c | c  c  c  c }
\hline
\rule{0pt}{8pt} & Method & F1 Score & $\log \theta$\\ \hline
& Base~\cite{unet} & \textbf{0.8644} & 7.492  \\ 
\hline
\multirow{2}{*}{\centering Test case 2} & Uniform multiplier & 0.8278 & 6.834 \\
& Layer-wise multiplier & 0.8273 & \textbf{5.925} \\
\hline
Pre-training compression & SqueezeNet~\cite{squeezenet} & 0.8267  & 7.049 \\
\hline
\multirow{4}{*}{\centering Post-training compression} & Taylor Pruning~\cite{nvidiapruning} & 0.8205 & 7.491 \\
& Dynamic Quantization~\cite{qbert} & 0.8249 & 7.492 \\
& CC-Net~\cite{suraj} $\alpha = 0.5$ & 0.8561 & 6.889 \\ 
& Importance Estimation Pruning~\cite{pavlo} & 0.8383 & 7.492 \\
\hline
\multirow{2}{*}{\centering Test case 2 + Neuron Merging~\cite{neuron_merging}} &  Uniform multiplier & 0.8405 & 6.834 \\
& Layer-wise multiplier & 0.8354 & \textbf{5.925} \\
\hline
\end{tabular}}
\end{center}
\label{tab:comparison}
\end{table*}

\begin{figure}[tb]
  \centering
 \includegraphics[width=0.8\textwidth]{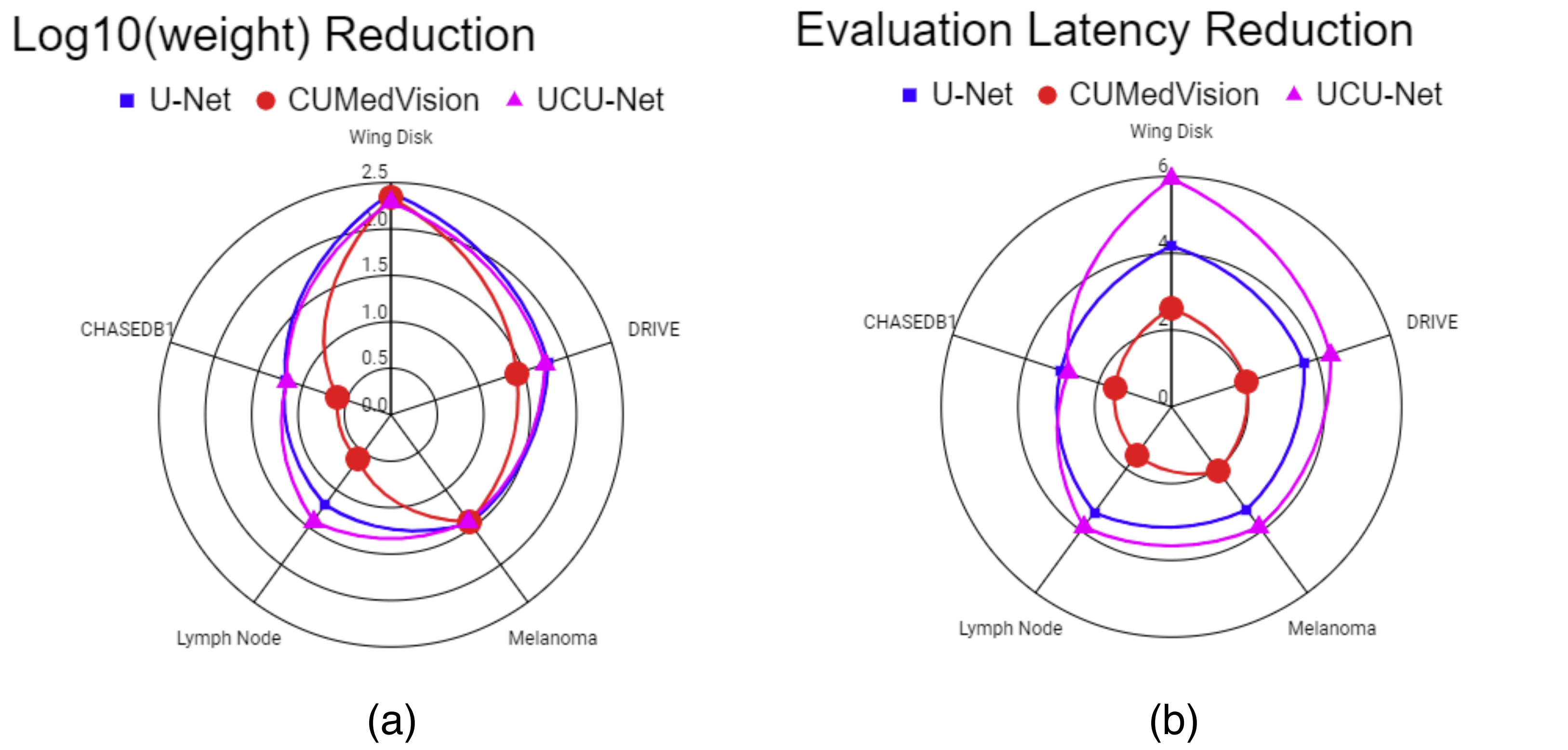}
\caption{Trainable weights and inference latency reduction achieved (on test case 1) for different datasets.}
\label{fig:gain_analysis}
\end{figure}

\subsection{Overall Gain}
The overall reduction (R = $\frac{base}{compressed}$) in trainable weights and evaluation latency for all five datasets for a 95\% accuracy threshold, approximated using uniform multiplier (as shown in Fig.~\ref{fig:unet-plots}, Fig.~\ref{fig:cunet-plots}, and Fig.~\ref{fig:ucu-plots}), is plotted in Fig.~\ref{fig:gain_analysis}(a) and Fig.~\ref{fig:gain_analysis}(b), respectively. Larger complexity results in less compression, indicating a higher requirement in trainable weights for extracting features. Our framework achieves best weight reduction ($\approx 247x$ on U-Net) and evaluation latency reduction ($\approx 6x$ on UCU-Net) in the case of the wing disk dataset. The least weight reduction ($\approx 4x$ on CUMedVision) and evaluation latency reduction ($\approx 2x$ on CUMedVision) are achieved for the most complex CHASE\_DB1 dataset.

\begin{figure}[tb]
  \centering
 \includegraphics[width=0.8\textwidth]{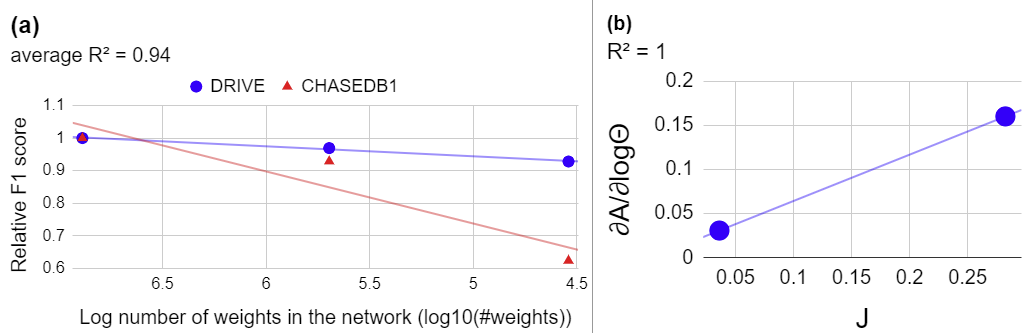}
\caption{Degree of degradation calculation using fewer data points.}
\label{fig:bottleneck}
\end{figure}

\subsection{Bottleneck Consideration}
One time determination of $\lambda$ and $\delta$ (i.e., the degree of degradation) for any CNN architecture is the bottleneck for our approach. Yet, once the degree of degradation is determined, significant reduction in training and evaluation time can be achieved for any dataset, trained and evaluated on the compressed network. We propose that the one time degree of degradation calculation can be performed with an acceptable level of accuracy by using two datasets with three $\alpha$ values ($\alpha \in \{1, 0.25, 0.0625\}$). We verify this by performing experiments on the CUMedVision network using the DRIVE and CHASE\_DB1 datasets, tracking the F1 score degradation with compression. Results thus obtained are shown in Fig.~\ref{fig:bottleneck}. Using fewer data points to determine the degree of degradation causes only 2.4\% change in the $\lambda$ value (new $\lambda$ = 0.525, new $\delta$ = 0.0116).

\section{Conclusions}
\label{sec:conclusion}
In this paper, we presented a new image complexity-guided deep learning based network compression approach for biomedical image segmentation. Instead of the usual practice of compressing CNN architectures after training, we focus on pre-training network compression, exploiting image complexity of the training data. Using the network's degree of degradation information, we showed that our approach is fast in predicting the compressed network's accuracy without training, and is effective in generating compressed networks. Our scheme accommodates practical applied design constraints for compressing CNNs for biomedical image segmentation by proposing fine-grain layer-wise multipliers. Such fine-grain control is capable of achieving better compression and better accuracy compared to uniform multiplier based compression techniques. Using five biomedical image segmentation datasets, we verified that our framework is capable of generating compressed networks, retaining up to $\approx 95\%$ of the full-sized network segmentation accuracy while utilizing significantly fewer trainable weights (in the range of $\approx 247x$ to $\approx 6x$ less).  

\section*{Acknowledgement}
This work was supported in part by the National Science Foundation under Grants CNS-1629914, CCF-1640081, and CCF-1617735, and by the Nanoelectronics Research Corporation, a wholly-owned subsidiary of the Semiconductor Research Corporation, through Extremely Energy Efficient Collective Electronics, an SRC-NRI Nanoelectronics Research Initiative under Research Task ID 2698.004 and 2698.005.

\bibliographystyle{ACM-Reference-Format}
\bibliography{sample-base}

\end{document}